\documentclass[11pt,a4paper]{article}
\usepackage{amsmath}
\usepackage{amsfonts}
\usepackage{amssymb}
\usepackage[left=1.5cm, right=1.5cm, top=1.50cm, bottom=1.50cm]{geometry}
\usepackage{natbib}
\usepackage{booktabs}
\usepackage{setspace}
\usepackage{subfigure}

\usepackage{graphicx}

\usepackage{fancyhdr}
\usepackage{rotating}
\usepackage{array}
\usepackage{color}
\usepackage{multirow}
\usepackage{float}
\usepackage{url}
\usepackage{subfigure}
\usepackage{hyperref}
\usepackage{abstract}
\usepackage{multicol}
\hypersetup{
unicode=false,          
pdftoolbar=true,        
pdfmenubar=true,        
pdffitwindow=false,     
pdftitle={Dayglow on Mars},    
pdfauthor={Sonal Kumar Jain},   
pdfsubject={Cameron band and UV doublet band},   
colorlinks=true,        
linkcolor=blue,         
citecolor=blue,        
filecolor=magenta,      
}

\newcommand{\car}{{$\mathrm{CO_2} $}}
\newcommand{\carp}{{$\mathrm{CO_2^+} $}}

\newcommand{\nt}{{$\mathrm{N_2}$}}
\newcommand{\dgr}{{$^\circ$}}

\title{\bf Model Calculation of N$_2$ Vegard-Kaplan band emissions in  Martian dayglow}
\date{}
\author{Sonal Kumar Jain\thanks{sonaljain.spl@gmail.com} and
Anil Bhardwaj\thanks{anil\_bhardwaj@vssc.gov.in; bhardwaj\_spl@yahoo.com} \\ 
Space Physics Laboratory,\\ 
Vikram Sarabhai Space Centre,\\ 
Trivandrum, India - 695022}

\begin{document}

\maketitle

\begin{abstract}
A model for N$_2$ Vegard-Kaplan band ($ A^3\Sigma^+_u - X^1\Sigma^+_g $) 
emissions in Martian dayglow has been developed to explain the 
recent observations made by the SPICAM ultraviolet spectrograph 
aboard Mars Express. Steady state photoelectron fluxes and  
volume excitation rates have been calculated using the 
Analytical Yield Spectra (AYS) technique. Since inter-state 
cascading is important for triplet states of N$_2$, the population 
of any given level of N$_2$ triplet states is calculated under statistical
equilibrium considering direct excitation, cascading, and 
quenching effects. Relative population of all vibrational levels of
each triplet state is calculated in the model.
Line of sight intensities and height-integrated overhead intensities have been
calculated for Vegard-Kaplan (VK), First Positive 
($ B^3\Pi_g - A^3\Sigma^+_u $), Second Positive 
($ C^3\Pi_u - B^3\Pi_g $), and Wu-Benesch ($W^3\Delta_u - B^3\Pi_g$) 
bands of N$_2$. A reduction in the N$_2$ density by a factor of 3 in the Mars 
Thermospheric General Circulation Model is required to obtain agreement 
between calculated limb profiles of VK (0-6) and SPICAM observation. 
Calculations are carried out to asses the impact of model parameters,
viz., electron impact cross sections, solar EUV flux, and 
model atmosphere, on the emission intensities. Constraining the  
N$_2$/CO$_2$ ratio by SPICAM observations, we suggest the  N$_2$/CO$_2$ ratios 
to be in the range 1.1 to 1.4\% at 120 km, 1.8 to  3.2\% at 
140 km, and 4 to 7\% at 170 km. During high solar activity the overhead
intensity of N$_2$ VK band emissions would be $ \sim $2.5 times higher than
that during low solar activity.
\end{abstract}

\begin{multicols}{2}
\section{Introduction}
Emissions from excited states of \nt\ have been studied 
extensively in the terrestrial airglow and aurora 
\citep[{\it e.g.},][]{Sharp71,conway85,Meier91,Morrill96,Broadfoot97}.
But the absence of any emission feature of \nt\ during Mariner observations
of Mars \citep{Barth71} surprised the planetary scientists
who attributed it to the low fractional abundance by volume
of molecular nitrogen on Mars \citep{Dalgarno70b}.
Earlier, \nt\ emissions on Mars were predicted by Fox and 
co-workers \citep{Fox77, Fox79},  who suggested that a high 
resolution UV spectrometer could detect the \nt\ UV emissions 
on Mars. \cite{Fox79} have predicted the intensity of 
various \nt\ triplet state emissions (Vegard-Kaplan, First positive, 
Second Positive, $W - B$), along with LBH band of \nt\ and First 
Negative band of N$_2^+$.
In the terrestrial atmosphere emission from Vegard-Kaplan (VK) bands are 
weak due to efficient quenching by atomic oxygen, but \car\ 
is not good at quenching VK bands \citep{Fox79,Dreyer74}, so the 
intensity of these band should be appreciable in Martian airglow.

Recent observations by SPICAM (Spectroscopy for Investigation of 
Characteristics of the Atmosphere of Mars) onboard Mars Express (MEX)
have, for the first time, observed \nt\ emissions in the 
dayglow of Mars \citep{Leblanc06,Leblanc07}. The
main emissions observed  are (0, 5), (0, 6), and (0, 7) bands  
of VK system, which originate from triplet $ A^3\Sigma^+_u $ 
state of excited \nt\ molecule. The overhead intensity of the
\nt\ VK (0, 6) band derived from the intensity observed by the SPICAM 
is found to be $ \sim $3 times smaller than the intensity
calculated by \cite{Fox79}.

There have been several measurements of electron impact cross sections of 
triplet states of \nt\ since \cite{Fox79} carried out their calculations.
With new cross sections and  updated molecular parameters 
(transition probability and Franck-Condon factor) a model of \nt\ 
dayglow emission on Mars is necessary for a better understanding 
of the recent SPICAM observations.
In the present work, a model has been developed to calculate 
the \nt\ dayglow emissions on Mars using the Analytical Yield 
Spectra approach. While calculating the emission of 
VK bands of \nt, cascading from the higher lying states and
quenching by atmospheric constituents are considered and 
the population of any given vibrational level of a
state is calculated under statistical equilibrium.
Height-integrated overhead intensities are reported for
major vibrational bands of \nt\ VK, First Positive 
($ B^3\Pi_g - A^3\Sigma^+_u $), Second Positive 
($ C^3\Pi_u - B^3\Pi_g $), and Wu-Benesch ($W^3\Delta_u - B^3\Pi_g$) bands.
Limb profiles of VK (0, 5), (0, 6), and (0, 7) bands are calculated. The
limb profile of VK (0, 6) band is compared with that reported by
the SPICAM observations. The present model has been used recently to estimate the
\nt\ triplet band intensities on the Venus \citep{Bhardwaj11b}.

\section{Vegard-Kaplan Band ($ A^3\Sigma^+_u \rightarrow X^1\Sigma^+_g $)}
\begin{figure}    
\centering 
\includegraphics[width=30pc]{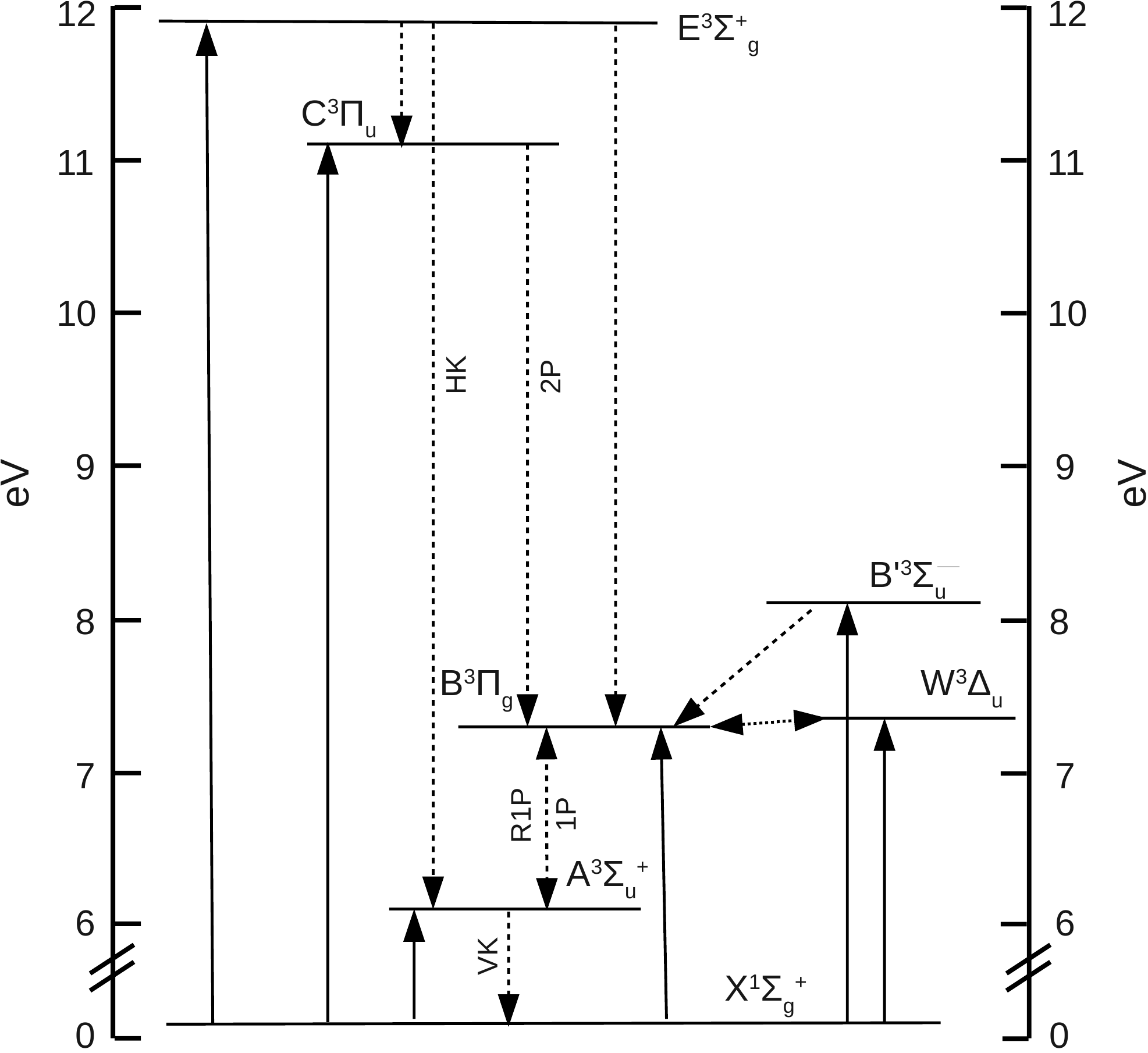}
\caption[]{Energy level diagram for the excitation of \nt\
triplet states and subsequent inter-state cascading processes. Solid arrows show
the excitation from ground state to higher states, and dashed arrows 
represent the transitions between different states (HK: Herman-Kaplan; 
1P: First Positive; R1P: Reverse First Positive; 2P: Second Positive; 
VK: Vegard-Kaplan band system). Excitation thresholds for all the triplet 
states are given Table~\ref{tab:table1}.}
\label{fig:n2el}
\end{figure}

Figure~\ref{fig:n2el} shows schematic diagram of \nt\ triplet states energy
level with excitation and subsequent cascading processes.
The transition from the ground state (X$^1\Sigma^+_g$) to the A$^3\Sigma^+_u$
state is dipole forbidden, so photoelectron impact is the primary excitation
source for this state. In addition to the direct excitation from the 
ground state, cascade from higher triplet states $C$, $B$, $W$, and 
$ B'$ are also important. All excitations of higher triplet states will
eventually cascade into the $ A^3\Sigma^+_u$ state \citep{Cartwright71,Cartwright78}.
The width and shape of VK bands are quite sensitive to
the rotational temperature, making them useful as a monitor of the neutral
temperature of the upper atmosphere \citep{Broadfoot97}.

All transitions between the triplet states of \nt\ and
the ground state are spin forbidden, therefore
excitation of these states is primarily due to the electron impact.
The higher lying states $C$, $W$, and  $ B'$
populate the $B$ state, which in turn radiates to the $A$ state.
Inter-system cascading $ B^3\Pi_g \rightleftharpoons A^3\Sigma^+_u $
and $ B^3\Pi_g \rightleftharpoons W^3\Delta_u $ is
important in populating the $ B $ state \citep{Cartwright71,Cartwright78}.

Direct excitation of the $ \nu'=0 $ vibrational level of the 
$ A^3\Sigma^+_u$ state by electron impact is extremely small, because
Frank-Condon factor to the $ \nu''=0 $ level of the ground electronic
state, q$_{00}$, is only $9.77 \times 10^{-4}$ \citep{Gilmore92, Piper93}.
Contributions to  $ \nu'=0 $ level of $A$ state come from the
higher states cascading.  We have also included
$ E \rightarrow B $, $ E\rightarrow C $, $ E\rightarrow A $, 
$ B\rightarrow W $, and reverse first positive $ A\rightarrow B $  cascading in 
our calculation. The effect of reverse first positive transition is important
in populating the lower vibrational levels of $B$ state, which 
in turn populate the lower vibrational levels of the $ A $ state
\citep{Sharp71,Cartwright71,Cartwright78}.
Thus, to calculate the production rate of any vibrational level of triplet
state of \nt, one must take into account direct excitation as well as
inter-state cascading effects.

\section{Model Input Parameters} 
\label{sec:mi}
The model atmosphere considering five gases (\car, CO, \nt, O, and O$_2$)
is taken from the Mars Thermospheric General Circulation Model (MTGCM) of 
\cite{Bougher90,Bougher99,Bougher00} for a solar longitude
of 180\dgr, latitude of 47.5\dgr N, and at 1200 LT; and is same as used in the
study of \cite{Shematovich08}.
The EUVAC model of \cite{Richards94} has been used to calculate the 37-bin
solar EUV flux for the day of observation, which is based on the F10.7 
and F10.7A (81-day average) solar index. The F10.7 flux as seen by 
Mars (by accounting for the Mars-Sun-Earth angle) is used to derive
the 37-bin solar EUV flux. The EUVAC solar spectrum thus obtained is then 
scaled for the heliocentric distance of Mars for the day,
considered in the present study. To assess the impact of solar 
EUV flux on model calculations, we have also used SOLAR2000 v.2.36 (S2K) model of
\cite{Tobiska04}.

\begin{figure}	 
\centering
\includegraphics[width=30pc]{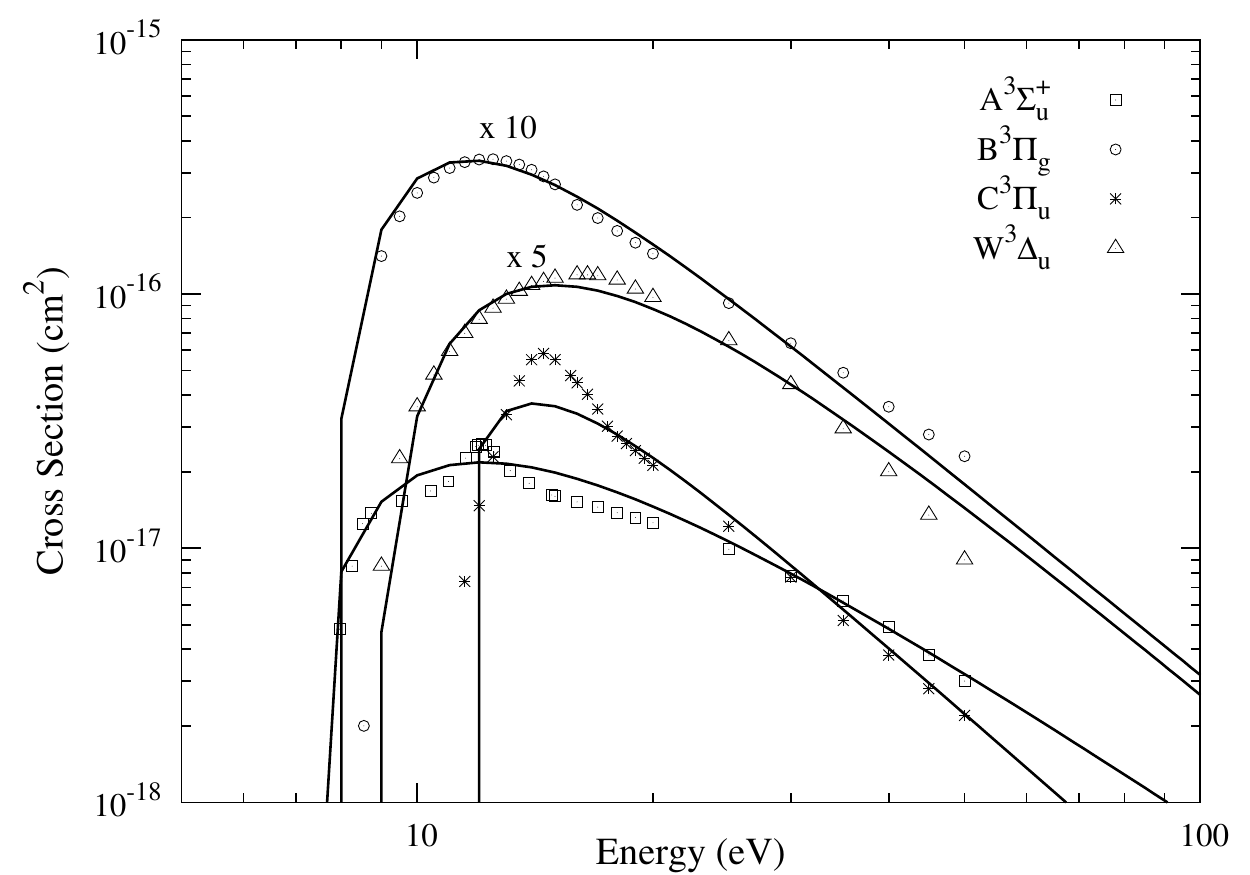}
\caption{The triple states cross sections due to electron impact on \nt. Symbols 
represent the values of \cite{Itikawa06} and the solid curve represents the 
analytical fits using equation~\ref{eq:jack}. Cross sections of $ B$ and $W$ have
been plotted after multiplying by a factor of 10 and 5, respectively.}
\label{fig:n2-xs}
\end{figure}

Photoionization and photoabsorption cross sections for the
gases considered in the present study are taken from \cite{Schunk00}.
The branching ratios for excited states of \carp, CO$^+$, N$_2^+$,
O$^+$, and O$_2^+$  have been 
taken from \cite{Avakyan98}. For calculating the intensity of a specific band
$ \nu'-\nu'' $, Franck-Condon factors and transition 
probabilities are required. For \nt\ these are taken from 
\cite{Gilmore92}.
Electron impact cross sections for \nt\ triplet excited states 
($ A,\, B,\, C, \, W, \,B',$ and $ E $) were  measured by 
\cite{Cartwright77} up to 50 eV. These cross sections were renormalized
later by \cite{Trajmar83} with the use of improved data on elastic
cross sections. More recently, \nt\ triplet state cross sections
have been measured by \cite{Campbell01} and \cite{Johnson05}.
\cite{Itikawa06} reviewed the cross sections of the \nt\ triplet excited states and
recommended the best values determined by \cite{Brunger03}.
We have taken the  \nt\ triplet states cross sections
from \cite{Itikawa06}, which have been fitted analytically
using equation \citep[cf.][]{Jackman77, Bhardwaj09}

\begin{equation}\label{eq:jack}
 \sigma(E)=\frac{(q_0F)}{W^2}\left[1-\left(\frac{W}{E}\right)^\alpha
 \right]^\beta \left[\frac{W}{E}\right]^\Omega,      
\end{equation}
where $q_0=4\pi a_0R^2$ and has the value $6.512\times10^{-14}$ eV$^2$
cm$^2$.
Table~\ref{tab:table1} shows the corresponding parameters.
Fig.~\ref{fig:n2-xs} shows the fitted cross sections of the \nt\ triplet
$ A,\, B,\, C$, and $W$ states along with the recommended cross sections
of \cite{Itikawa06}. For other gases electron impact cross sections have
been taken from \cite{Jackman77},  except for \car, which are  from	
\cite{Bhardwaj09}.

We have run our model for the Mars Express observation on 16 Dec. 2004
(Sun-Mars distance = 1.59 AU, and F10.7 at Mars = 35.6),
taking solar zenith angle as 45\dgr, solar EUV flux from the EUVAC model, and
MTGCM model atmosphere. Hereafter we
refer it as the ``standard case''. We have also studied the effects of various
input parameters (like solar EUV flux, \nt\ triplet state cross sections,
model atmosphere, solar cycle) on the emission intensity, which are discussed
in Section~\ref{sec:ipi}.

\section{Model Calculation}
\label{sec:mc}
\subsection{Photoelectron Production Rate}
\label{subsec:mcppr}
\begin{figure}     
\centering
\includegraphics[width=30pc]{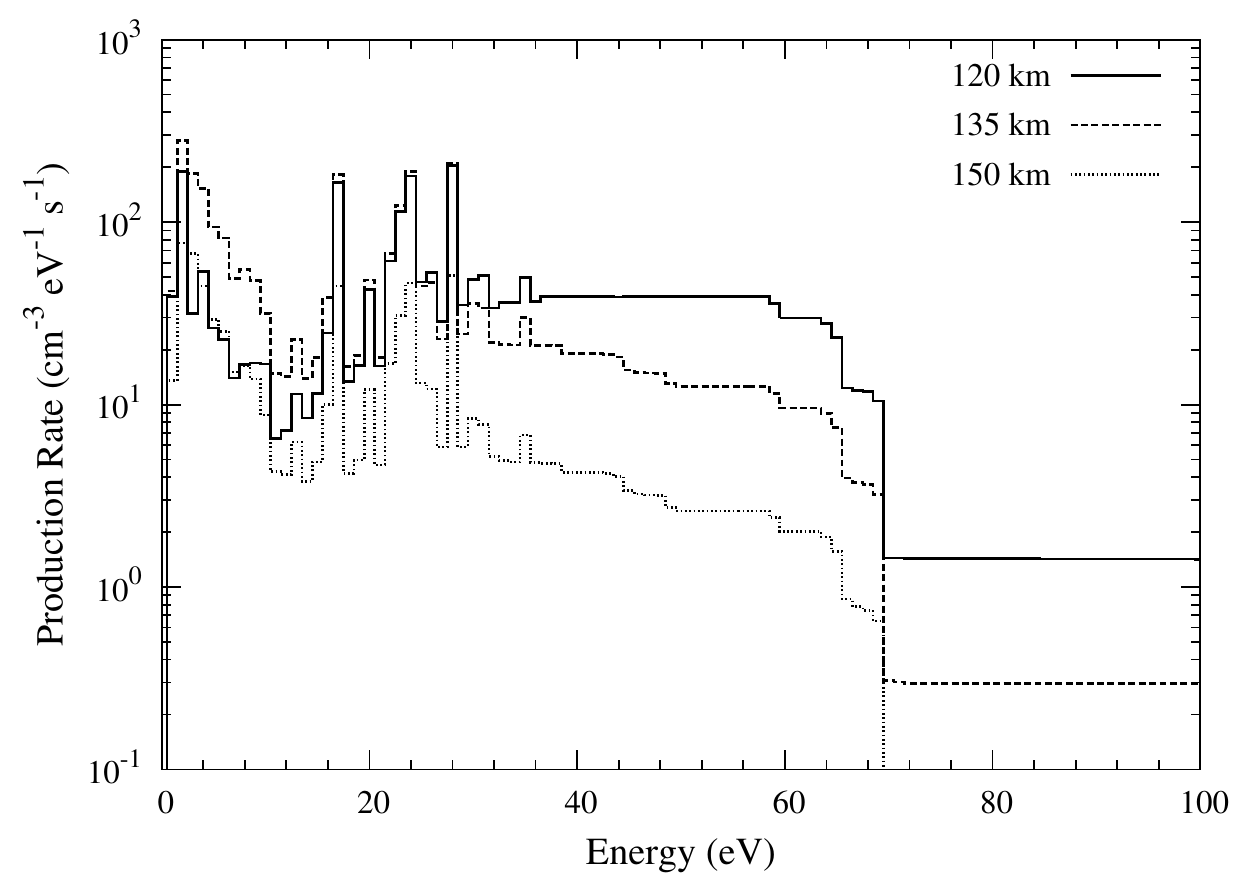}
\caption{Primary photoelectron energy distribution at three different altitudes for the standard case.}
\label{fig:per}
\end{figure}

Primary photoelectron production rate is calculated using 

\begin{equation}
Q(Z,E)=  \sum_l n_l(Z) \sum_{j,\lambda}\sigma_l^I(j,\lambda) I(Z,\lambda) \
\delta \left(\frac{hc}{\lambda}-E-W_{jl}\right)
\end{equation}
\begin{equation}
I(Z,\lambda)=I(\infty,\lambda)\ exp\left[-\sec(\chi)\sum_{l} \sigma_l^{A}
(\lambda)\int_Z^{\infty}n_l(Z^{'})dZ^{'}\right]
\end{equation}
where $\sigma_l^A$ and $\sigma_l^I (j,\lambda)$  are the total photoabsorption 
cross section and the photoionization cross section of the $j$th ion state of
the constituent $l$ at wavelength $\lambda$, respectively;
$I(\infty,\lambda)$ is the unattenuated solar flux at wavelength $\lambda$, 
$n_l$ is the neutral density of constituent $l$ at altitude Z; $\sec$($\chi$) is 
the Chapman function, $\chi$ is the solar zenith angle (SZA); 
$\delta(hc/\lambda-E-W_{jl})$ is the delta function, 
in which $hc/\lambda$ is the incident photon energy, W$_{jl}$ is 
the ionization potential of $j$th ion state of the $l$th constituent, and $E$ is 
the energy of ejected electron. We have used $\sec(\chi)$ in place of 
ch($\chi$), which is valid for $\chi$ values upto 80\dgr. Figure~\ref{fig:per}
shows the primary photoelectron energy spectrum  at three different altitudes. There is a
sharp peak at 27 eV due to the ionization of \car\ in the ground state
by the He II solar Lyman $ \alpha $ line at 303.78 \AA.
The peaks at 21 and 23 eV are due to ionization of \car\ in the A$^2\Pi_u$ 
and B$^2\Sigma_u^+$ states of \carp, respectively, by the 303.78 \AA\ solar photons.
The individual peaks structure shown in the figure are
different from that of \cite{Mantas79}, which is due to revisions
in the branching ratios \citep{Avakyan98}  used in the present study.

\subsection{Photoelectron Flux}
To calculate the photoelectron flux we have adopted the Analytical Yield 
Spectra (AYS) technique \citep[cf.][]{Singhal84,Bhardwaj90a,Bhardwaj90b,Bhardwaj96,
Singhal91,Bhardwaj99a,Bhardwaj03,Bhardwaj99d,Bhardwaj99b}.
The AYS is the analytical representation of numerical 
yield spectra obtained using the Monte Carlo model \citep[cf.][]{Singhal80,
Bhardwaj99d,Bhardwaj99b,Bhardwaj09}.
Recently, the AYS model for electron degradation in \car\ 
has been developed by \cite{Bhardwaj09}. Further details of the AYS technique are
given in \cite{Bhardwaj99d}, \cite{Bhardwaj09}, and references therein.  Using AYS the 
photoelectron flux has been calculated as \citep[e.g.][]{Singhal84,Bhardwaj99b}

\begin{equation}\label{eq:a}
\phi(Z,E)=\int_{W_{kl}}^{100} \frac{Q(Z,E) U(E,E_0)}
{{\displaystyle\sum_{l}} n_l(Z)\sigma_{lT}(E)} \ dE_0
\end{equation}
where $\sigma_{lT}(E)$ is the total inelastic cross section for the $l$th gas, 
$n_l$ is its density, and $U(E,E_0)$ is the two-dimensional AYS,
which embodies the non-spatial information of degradation process. It 
represents the equilibrium number of electrons per unit energy at an
energy $E$ resulting from the local energy degradation of an incident
electron of energy $E_0$. For the \car\ gas it is given as \citep{Bhardwaj09}

\begin{equation}\label{eq:b}
      U(E,E_0)=A_1E_k^s+A_2(E_k^{1-t}/\epsilon^{3/2 +r})+ \frac{E_0B_0e^{x}/B_1}{(1+e^{x})^2}
\end{equation}
Here $E_k=E_0/1000$, $\epsilon=E/I$ ($I$ is the lowest ionization
threshold), and $x=(E-B_2)/B_1$. $A_1=0.027,\ A_2=1.20,\ t=0,\ 
r=0$, $s=-0.0536$, $B_0=10.095$, $B_1=5.5$, 
and $B_2=0.9$ are the best fit parameters.

For other gases, viz., O$_2$, \nt, O, and CO, we have used the AYS
given in \cite{Singhal80}

\begin{equation}\label{eq:d}
      U(E,E_0)=C_0+C_1(E_k+K)/[(E-M)^2+L^2].
\end{equation}
Here $C_0$, $C_1$, $K$, $M$, and $L$ are the
fitted parameters which are independent of the energy, and 
whose values are given by \cite{Singhal80}.

\begin{figure}      
\centering
\includegraphics[width=30pc]{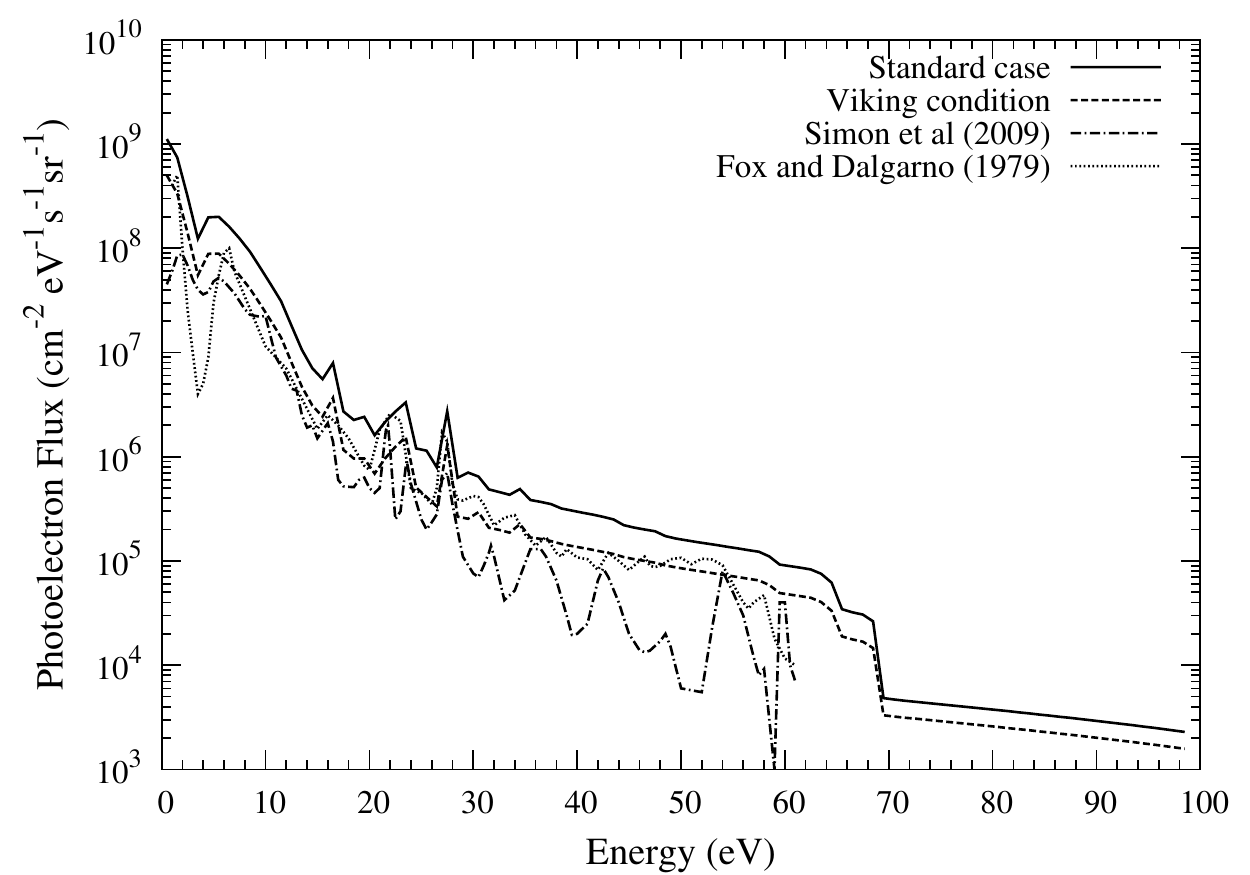}
\caption{Model steady-state photoelectron flux calculated at 130 km for standard case 
and for Viking condition. Flux calculated by \cite{Simon09} and 
\cite{Fox79} at 130 km are also shown for comparison.}
\label{fig:pef}
\end{figure}

The calculated photoelectron flux at 130 km altitude is shown 
in Figure~\ref{fig:pef} for the standard case as well as for 
conditions similar to those of Viking 1 (see Section~\ref{subsec:loi}).
The photoelectron flux calculated by \cite{Simon09} and \cite{Fox79} are also shown 
in Figure~\ref{fig:pef} at same altitude. Overall important peak structures 
are similar in all the three calculated fluxes, \textit{e.g.}, the peak at 27 eV and 
broad peak at 21-23 eV. A sharp dip at around 3 eV is 
prominent in all three photoelectron fluxes, which is due to large 
vibrational cross sections at 3.8 eV for electron impact on \car.
The calculated fluxes decrease exponentially with increasing energy.
The	sudden decrease in the photoelectron flux at higher energies
is due to the presence of these features in the primary photoelectron energy
spectrum (cf. Figure~\ref{fig:per}).


\section{Results and discussion}
\subsection{Volume excitation rates}
\begin{figure}    
\centering 
\includegraphics[width=30pc]{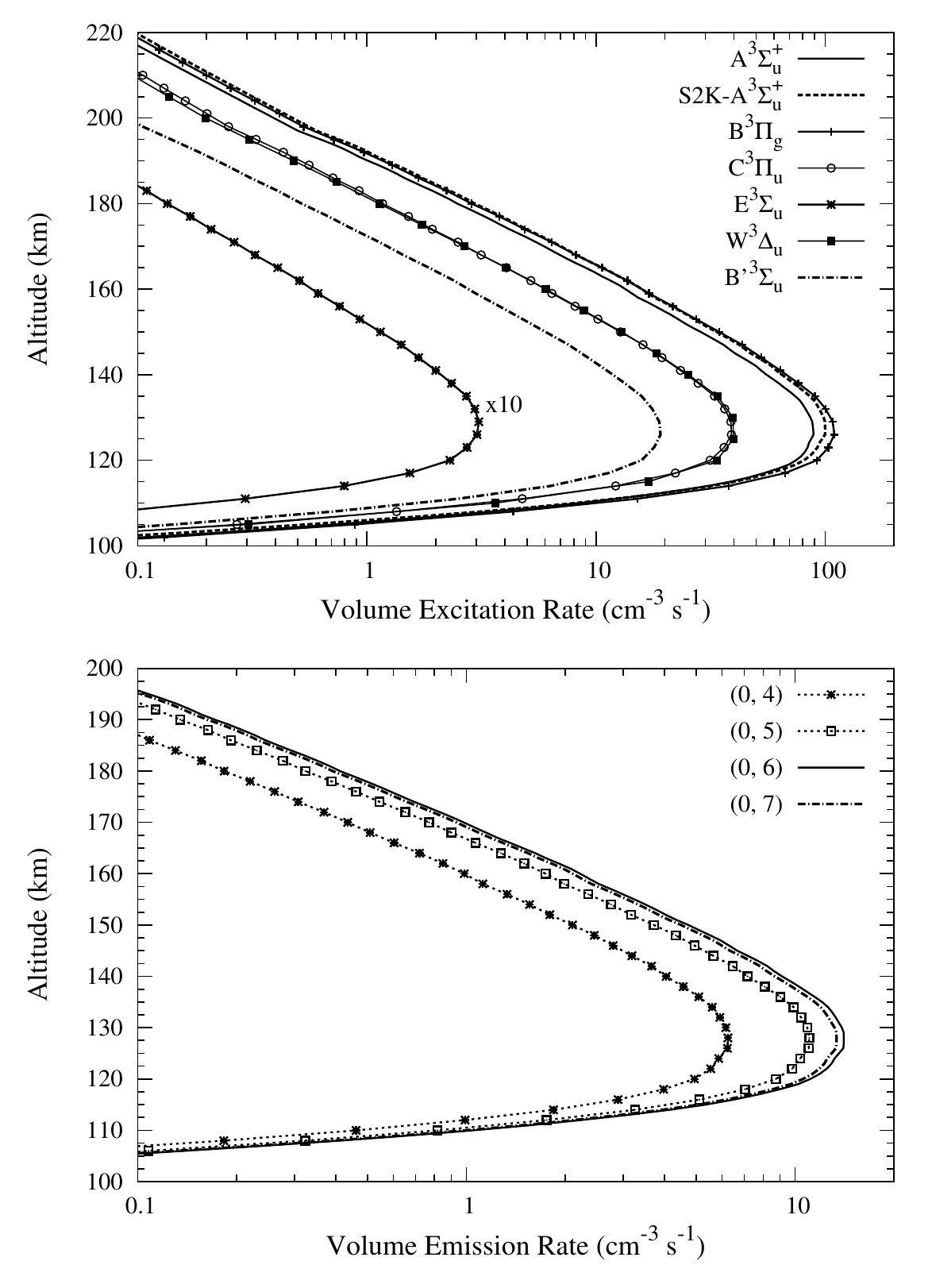}
\caption[]{(Upper panel) The volume excitation rates  of various 
triplet states of \nt\ by direct electron impact
excitation for the standard case. Dashed curve shows the excitation rate of
$A$ state calculated using S2K model.
The excitation rate of the $E$ state
has been multiplied by a factor of 10. (Bottom panel)
The volume emission rates of 
the VK (0, 4), (0, 5), (0, 6), and (0, 7) bands.}
\label{fig:n2ver}
\end{figure}

We have calculated volume excitation rate V$_{il}(Z,E)$ for the $i$th state
of the $l$th gas at altitude $Z$ and energy $E$ using the equation
\citep{Singhal91,Bhardwaj99a, Bhardwaj03,Bhardwaj99b}

\begin{equation}\label{eq:e}
V_{il}(Z, E) = n_l(Z) \int _{E_{th}}^{E} \phi(Z, E) \sigma_{il}(E) dE,
\end{equation}
where $n_l(Z)$ is the density of the $l$th gas at altitude $ Z $ and 
$ \sigma_{il}(E)$ is the electron impact cross section for the $i$th state
of the $l$th gas, for which the threshold is $E_{th}$. Figure~\ref{fig:n2ver} (upper panel)
shows the volume excitation rates of the \nt\ triplet states ($ A,\, B,\, C, \, W,
\,B',$ and $ E $) excited by photoelectron impact. The altitude
of peak production for all states is $ \sim $126 km for the standard case.
The volume excitation rate of \nt(A) state calculated using the S2K solar flux
model is also shown in the upper panel of Figure~\ref{fig:n2ver}. The peak of
excitation rate occurs at the same altitude for both solar EUV flux models but
the magnitude of excitation rate is slightly higher when the S2K model is used.
More discussion about the effect of solar EUV flux model on emission intensities
is given in Section~\ref{subsec:flx-model}.

To calculate the contribution of cascading from higher triplet states
and interstate cascading between different states, we solve the
equations for statistical equilibrium based on the formulation
of \cite{Cartwright78} and assumed that only excitation from the lowest 
vibrational level of the electronic ground state is important.  At a
specified altitude, for a  vibrational level $ \nu $  of a state $ \alpha $,
the population is determined using statistical equilibrium

\begin{equation}\label{eq:sta}
V^\alpha q_{0\nu}  + \sum\limits_{\beta}\sum\limits_{s}A^{\beta\alpha}_{s\nu}\, n^\beta_s
= \{ K^\alpha_{q\nu} + \sum\limits_{\gamma}\sum\limits_{r}A^{\alpha\gamma}_{\nu r} \}n^\alpha_{\nu}
\end{equation}
where
\begin{tabbing}
$ \alpha, \beta, \gamma $ \quad \= electronic states \kill
$ V^\alpha $ \> electron impact volume excitation rate \\ 
			\>	(cm$^{-3}$\ s$^{-1}$) of state $\alpha$;\\
$q_{0\nu}$   \> Franck-Condon factor for the  excitation \\
             \> from ground level to $\nu$ level of state $ \alpha $; \\
$A^{\beta\alpha}_{s\nu}$ \> transition probability (s$^{-1}$) from 
                         state  \\
               \>    $ \beta(s)$ to $ \alpha(\nu)$; \\	
$K^\alpha_{q\nu}$ \>  total electronic quenching frequency  \\
                 \> (s$^{-1}$) of level $\nu$ of state $ \alpha $ by the \\
                  \> all gases defined as: $\sum\limits_{l} K_{q(l)\nu}^\alpha \times n_{l} $; 
                     where, \\
                  \> $K_{q(l)\nu}^\alpha$ is the quenching rate coefficient  \\
                   \> of level $ \nu $ of $ \alpha $ by gas $ l $  of density $ n_l $;\\
$ A^{\alpha\gamma}_{\nu r} $ \> transition from level $ \nu $ of state $ \alpha $ 
                              to \\
                       \> vibrational  level $r$ of state $ \gamma $;\\
$ n $ \> density (cm$^{-3}$);\\
$ \alpha, \beta, \gamma $ \> electronic states;\\
$ s, r $ \> source and sink vibrational levels, \\
		\> respectively.
\end{tabbing}

While calculating the cascading from  $ C $ state, 
we have taken predissociation also
into account.  The $ C $ state predissociates approximately 
half the time (this is an average value for all 
vibrational levels of the $ C $ state; 0 and 1 levels do not
predissociate at all) \citep[cf.][]{Daniell86}. In the terrestrial
thermosphere, the \nt(A) state is effectively quenched by atomic oxygen.
In the case of Mars the main constituent \car\ does not quench \nt(A) level 
that efficiently, but still there will be some collisional deactivation
by other atmospheric constituents of Mars. The electronic quenching rates for 
vibrational levels of \nt\ triplet states by O, O$_2$, and \nt\ 
are adopted from \cite{Morrill96} and \cite{Cartwright78} and by \car\ and CO are 
taken from \cite{Dreyer74}.

\begin{figure}      
\centering
\includegraphics[width=30pc]{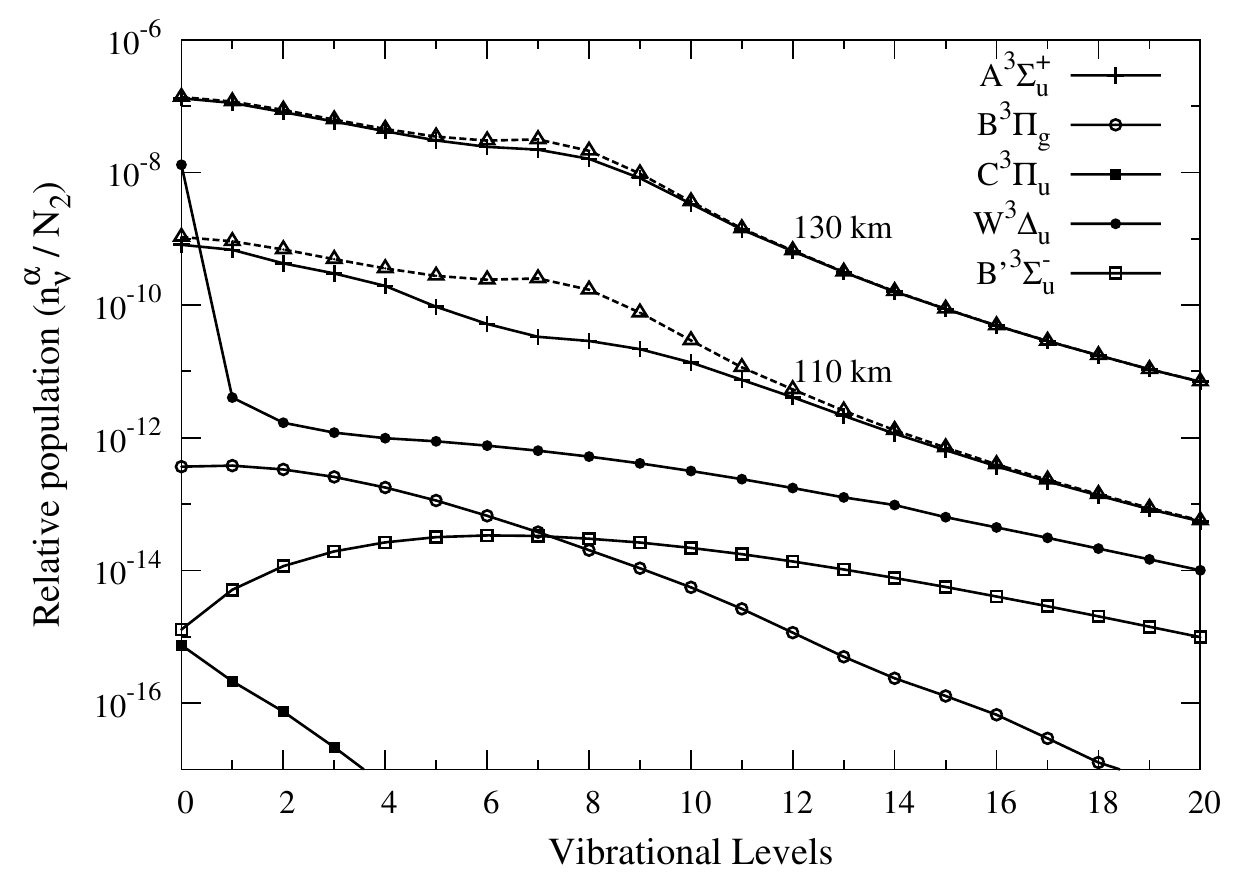}
\caption{The relative populations of vibrational levels of 
different triplet states of \nt\ with respect 
to the \nt\ density at 130 km. Dashed line with triangle shows
the relative vibrational populations of  $A$ at 110
and 130 km, respectively, without considering the quenching.}
\label{fig:vibpop}
\end{figure}

\begin{figure}     
\centering
\includegraphics[width=30pc]{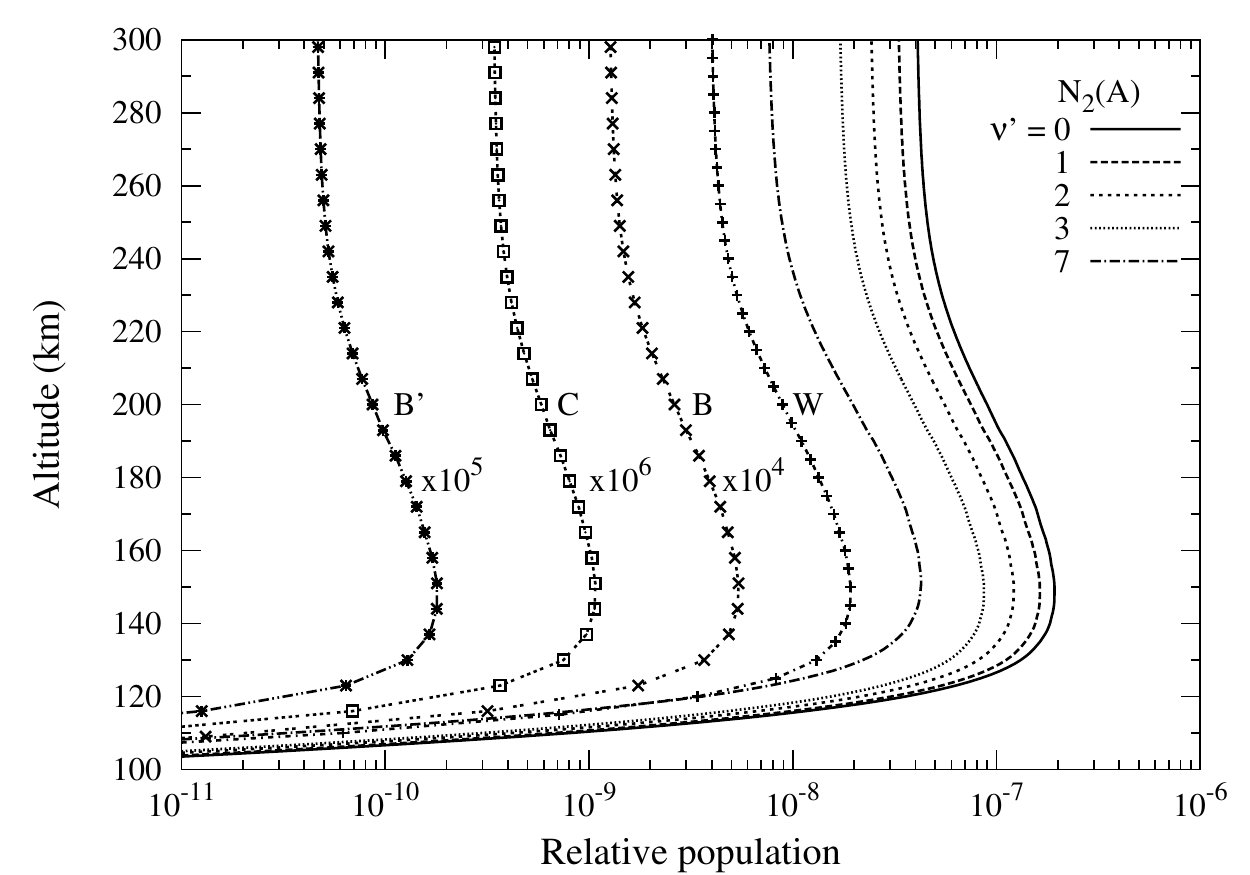}
\caption{Altitude profiles of the relative populations of selected vibrational 
levels of the \nt(A) state, and 0 level of the $ B$, $B'$, $C$, and $W$ states
with respect to those of \nt(X). Population of $B, B'$, and $C$ have 
been plotted after multiplying by a factor of $10^4, 10^5$, and $10^6$,
respectively.}
\label{fig:fpop-A}
\end{figure}

Figure~\ref{fig:vibpop} shows the population of different 
vibrational levels of triplet states of \nt\ relative 
to the ground state at 130 km. The relative population of \nt(A) at 110 km is 
also shown in the figure.
Our calculated relative vibrational populations agree well 
with the earlier calculations \citep{Morrill96,Cartwright78}.
To show the effect of quenching the relative
vibrational populations of \nt(A) state calculated without
quenching at 110 and 130 km are also shown in 
Figure~\ref{fig:vibpop}.
The quenching does affect the vibrational population of
\nt(A) state mainly for vibrational levels between 5 and 10
at lower altitudes ($<$130 km), as the altitude increases the
effect of quenching decreases. Figure~\ref{fig:fpop-A} shows the 
steady state fractional population altitude profiles of a few vibrational 
levels of $A$ state and $ \nu' = 0 $ level of $ B$, $C$, $W$,
and $ B' $ excited states of \nt.

After calculating the steady state density of 
different vibrational levels of excited states of 
\nt, the volume emission rate $ V_{\nu'\nu''}^{\alpha\beta} $ 
of a vibration band $\nu' \rightarrow \nu''$ can be obtained using

\begin{equation}\label{eq:g}
V_{\nu'\nu''}^{\alpha\beta} = n_{\nu'}^\alpha \times 
                            A_{\nu'\nu''}^{\alpha\beta} \quad (cm^{-3}\ s^{-1})
\end{equation}
where  $n_{\nu'}^\alpha$ is the density of vibrational 
level $\nu'$ of state $ \alpha $, and $A_{\nu'\nu''}^{\alpha\beta}$
is the transition probability (s$^{-1}$) for the transition from the
$\nu'$ level of the $\alpha$ state to the $\nu''$ level of the $\beta$
state. Figure~\ref{fig:n2ver} (bottom panel) shows the volume emission rates for the
VK (0, 4), (0, 5), (0, 6), and (0, 7) bands. We have integrated the volume
emission rates over the altitudes $80 - 400$ km to obtain
the overhead intensity for various VK bands of \nt, which are tabulated 
in Table~\ref{tab:tableA} (standard case). Table~\ref{tab:n2-oi}
shows the calculated height-integrated overhead intensities for a few of
the prominent bands of First Positive ($B \rightarrow A$), Second Positive
($ C\rightarrow B $), and Wu-Benesch ($ W \rightarrow $ B) emissions.

\subsection{Line of sight intensity}
\label{subsec:loi}
For comparison of the calculated intensity with SPICAM observation we have 
integrated the calculated emission rate along the line of sight and 
expressed the results in kR (1 Rayleigh = 10$^6$ photon 
cm$^{-2}$ s$^{-1}$)

\begin{equation}\label{eq:h}
I = \int \mathrm{V}(r)dr,
\end{equation}
where V(r) is the volume emission rate (in cm$^{-3}$ s$^{-1}$) for a particular 
emission, calculated using
equation~(\ref{eq:g}) and $r$ is abscissa along the horizontal line of sight. 
The upper limit of the atmosphere in our model is taken as 400 km. While 
calculating limb intensity we assume that the emission rate is constant 
along local longitude/latitude. For the emissions considered in 
the present study, the effect of absorption in the atmosphere is 
found to be negligible. As mentioned earlier \citep[cf.][]{Leblanc07}, the
main \nt\ emission features observed by SPICAM are (0, 5) and (0, 6)
transitions of the Vegard-Kaplan ($A^3 \Sigma_u^+ - X^1\Sigma^+_g$) band.
\cite{Leblanc06} also reported the detection of VK (0, 7) band, but it was
characterized by a large uncertainty because it falls between two intense
emissions at 289 nm and 297.2 nm of \carp\ UV doublet and oxygen line emission,
respectively. Otherwise, as shown in Table~\ref{tab:tableA}, VK(0, 7) band
would have been more intense than the (0, 5) band. The ratio between calculated
intensity of the VK (0, 6) and (0, 5) bands is 1.3, which is in good agreement with
the results of \cite{Leblanc07} and \cite{Fox79}.

\begin{figure}    
\centering 
\includegraphics[width=30pc]{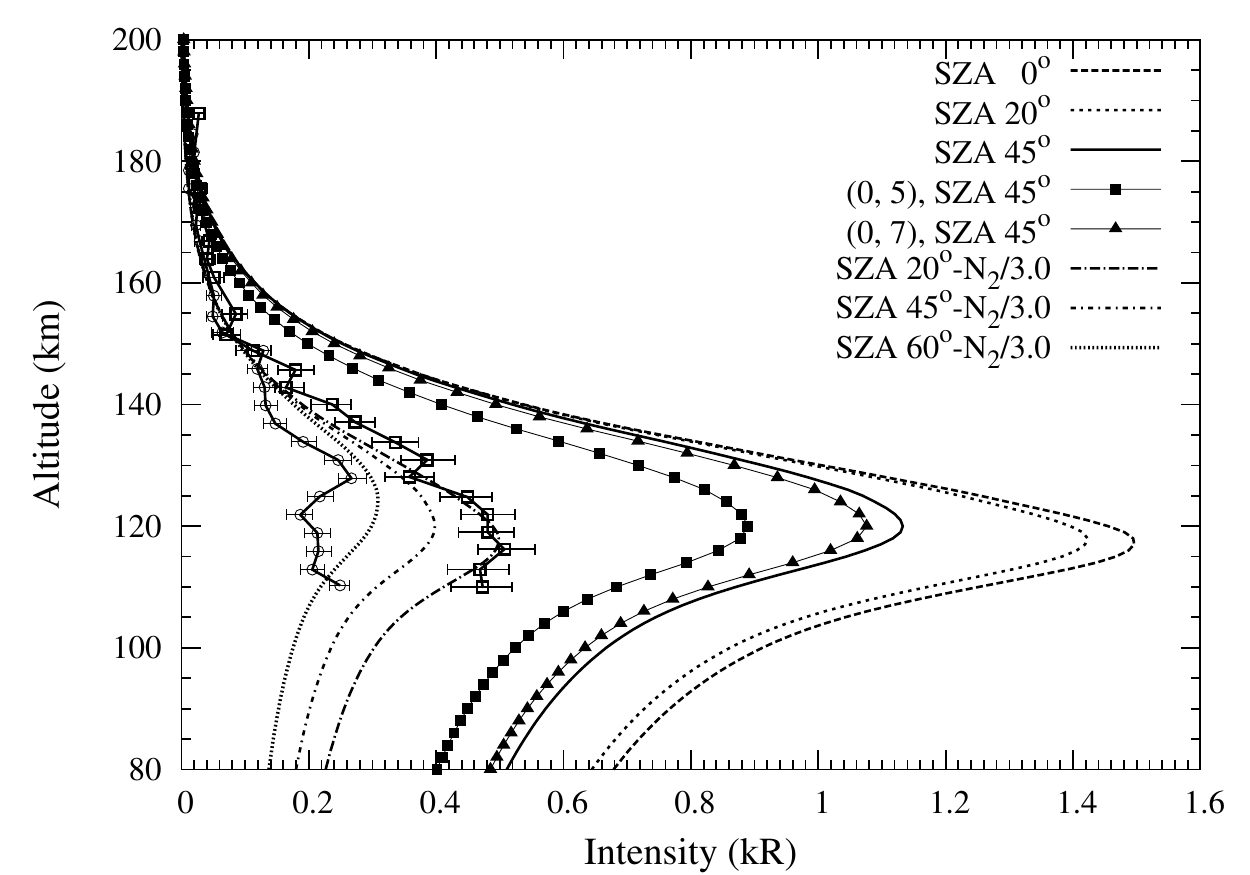}
\caption{Calculated limb intensity of the
\nt\ VK (0, 6) band at different solar zenith angles and for the VK (0, 5) and
(0, 7) at SZA = 45\dgr\ for the  standard case.
Lines with symbols (open squares, SZA = 8\dgr --36\dgr; open circles, 
SZA = 36\dgr -- 64\dgr) represent the averaged observed
value of the VK (0, 6) band for solar longitude (Ls) between 
100\dgr\ and 171\dgr\ taken from \cite{Leblanc07}. The calculated intensities,
when the \nt\ density is reduced by a factor of 3, are also shown.}
\label{fig:n2limb}
\end{figure}

Figure~\ref{fig:n2limb} shows the limb profiles of the VK
(0, 6) band at different solar zenith angles along with 
the SPICAM observed profiles averaged over the solar longitude 
L$_\mathrm{S}$ 100\dgr --171\dgr\, and SZA 
8\dgr--36\dgr\ and  36\dgr--64\dgr, taken from \cite{Leblanc07}.
The effect of SZA on the calculated profiles is clearly visible
in Figure~\ref{fig:n2limb}; the peak of the altitude profile rises
while the intensity  decreases with increasing SZA.
The limb profiles of the VK (0, 5) and (0, 6) bands at SZA=45\dgr\ are
also plotted in Figure~\ref{fig:n2limb}.
For the standard case (SZA=45\dgr), the peak intensities of the VK (0, 5),
(0, 6), and (0, 7) bands are $\sim$0.9, 1.1, and 1 kR, respectively, at 120 km.
For SZA values of 20\dgr\ and 60\dgr,  the \nt\ VK (0, 6) band peaks at
118 and 124 km with a value of 1.4 and 0.9 kR, respectively.

The shape of calculated and observed limb intensities are in
agreement with each other but the magnitude of calculated intensities 
are larger by a factor of $ \sim $3 at SZA = 20\dgr.
This difference could be due to the larger
abundance of \nt\ in the model atmosphere used in the present 
study. Other factors can also affect the calculated intensities, 
but their combined uncertainties also cannot account for the
difference by a factor of 3 in the calculated and observed 
intensities (effect of other input parameters, viz., electron 
impact cross section, and solar EUV flux model is described 
in Sections~\ref{subsec:ele-xs} and \ref{subsec:flx-model}, 
respectively).
Figure~\ref{fig:n2limb} also shows the computed limb intensity 
of the VK (0, 6) emission at SZA 20\dgr, 45\dgr, and 60\dgr\ obtained
after reducing the density of \nt\ 
by a factor of 3, which compares favourably  in both shape and 
magnitude with the observed emission. The \nt/\car\ ratio, after
reducing the \nt\ density by a factor of 3 is  to 0.9, 2.1, and 7.1\%
at altitudes of 120, 140, and 170 km, respectively. The calculated overhead 
intensities of VK bands after reducing \nt\ density by a factor of 3
(for the standard case) are depicted in column 3 of Table~\ref{tab:tableA}.
It may however be noted that the observed limb profiles
\citep{Leblanc07} are averaged over several days of 
observation (L$_s$=101\dgr-171\dgr) and range of SZA values, while the 
model profile is for a single day (16 Dec. 2004) at L$_s$ = 130\dgr\
and SZA = 20\dgr.

We have also calculated the nadir intensity for the condition similar 
to that of Viking landing (Sun-Mars distance = 1.65 AU and F10.7 = 68).
The model atmosphere was taken from \cite{Fox04} for the low solar
activity condition and a SZA of 45\dgr. For the VK
(0, 6) band our calculated intensity is 26 R, which is consistent 
with results (20 R) of \cite{Fox79} for the similar condition. The 
minor difference may be due to the updated cross sections and
transition probabilities. For the condition similar to that of 
Viking, \cite{Leblanc07} have measured an intensity of 
$ \sim $180 R for the VK (0, 6) band, which corresponds to a nadir intensity of 
$ \sim $6 R. The measured value is about 4 times smaller than our calculated
intensity. Such a difference by a factor of 4 between observed and 
calculated intensities might be due to the higher density of \nt\ 
taken in our model atmosphere. \cite{Leblanc07} mentioned 
that difference by factor of 3 between the estimated and 
nadir intensity calculated by \cite{Fox79} could have been due to the
larger \nt/\car\ ratio in the model atmosphere of \cite{Fox79}.
\cite{Leblanc07} have suggested that ratio of the integrated column densities 
of \nt\ and \car\ between 120 and 170 km, that is the 
mixing ratio between \nt\ and \car\ for a uniformly mixed atmosphere,
would be 0.9\% for an overhead intensity of 6 R. 
For the same altitude range, the ratio of \nt/\car\ density is 3.5\%
and 3.7\%  in model atmosphere used in the work of \cite{Fox04} and 
Bougher's MTGCM, respectively, which is a factor of 4 higher than that
suggested by \cite{Leblanc07}.

To summarize, the above results indicate that the \nt\ density in the MTGCM 
atmosphere, as well as in the model atmosphere of \cite{Fox04}, has to be reduced by
a factor of $ \sim $3 to obtain agreement between the SPICAM observation and the
calculated intensity.

\subsection{Variation with Solar Zenith Angle and Solar 10.7 flux}
\begin{figure}    
\centering 
\includegraphics[width=30pc]{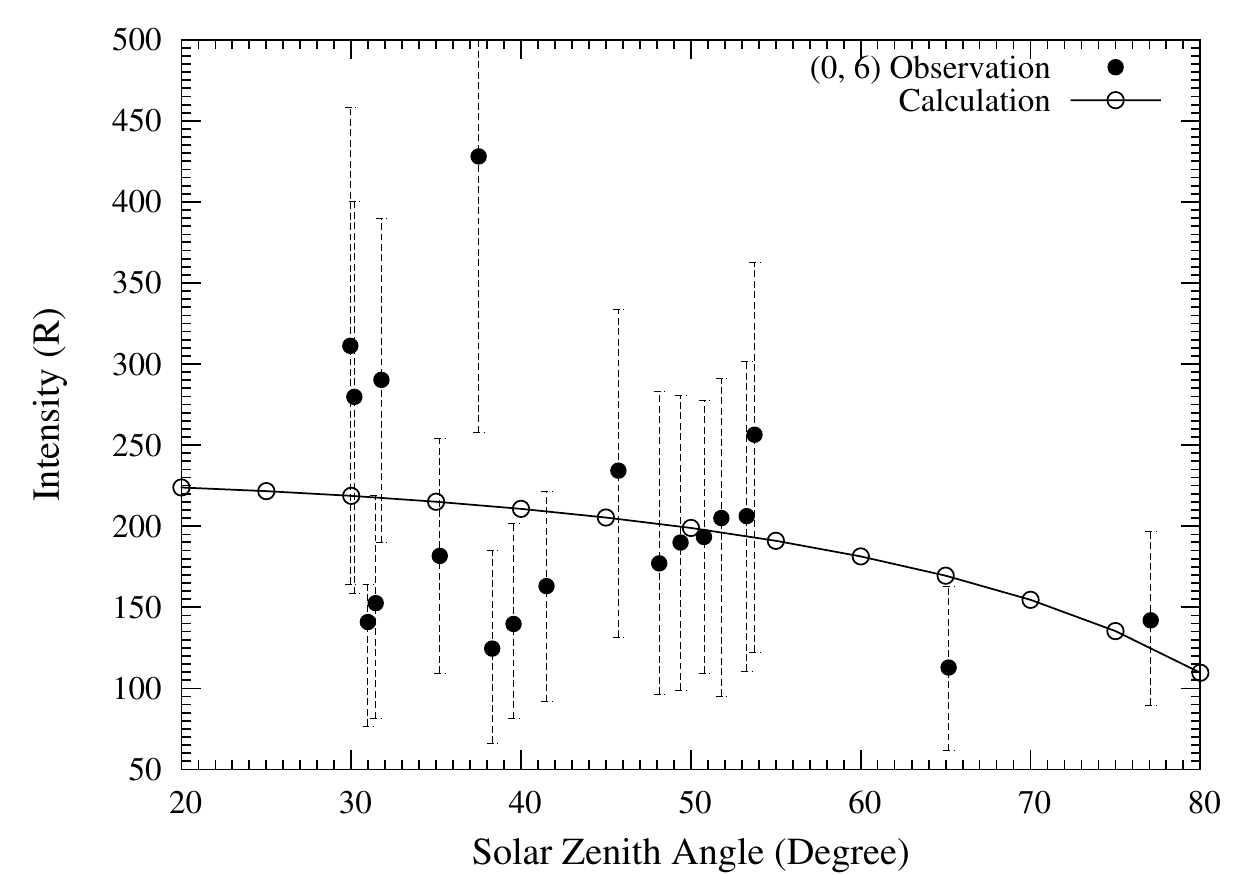}
\caption{The variation of the intensity of  the \nt\ VK (0, 6) emission with 
respect to solar zenith angle. The observed intensity of the VK (0, 6) band
is taken from Figure 2 of \cite{Leblanc07}. The calculated intensity is
averaged-value between 120 and 170 km for the standard case
with \nt\ density in the atmosphere reduced by a factor of 3.}
\label{fig:n2sza}
\end{figure}

Figure~\ref{fig:n2sza} shows the variation of the VK (0, 6) band 
intensity, averaged between 120 and 170 km, with SZA and 
its comparison with SPICAM observations. 
Calculated intensities are for standard case obtained after 
reducing the \nt\ density profile in the MTGCM atmosphere by a 
factor of 3 (see discussion in the previous section). Model
intensity shows a cosine SZA dependence, with larger 
attenuation of solar EUV flux at higher SZA, resulting in decrease in
the intensity at higher SZA. Calculated intensities are in agreement
with the observed values, within observational and model uncertainties.

\begin{figure}    
\centering 
\includegraphics[width=30pc]{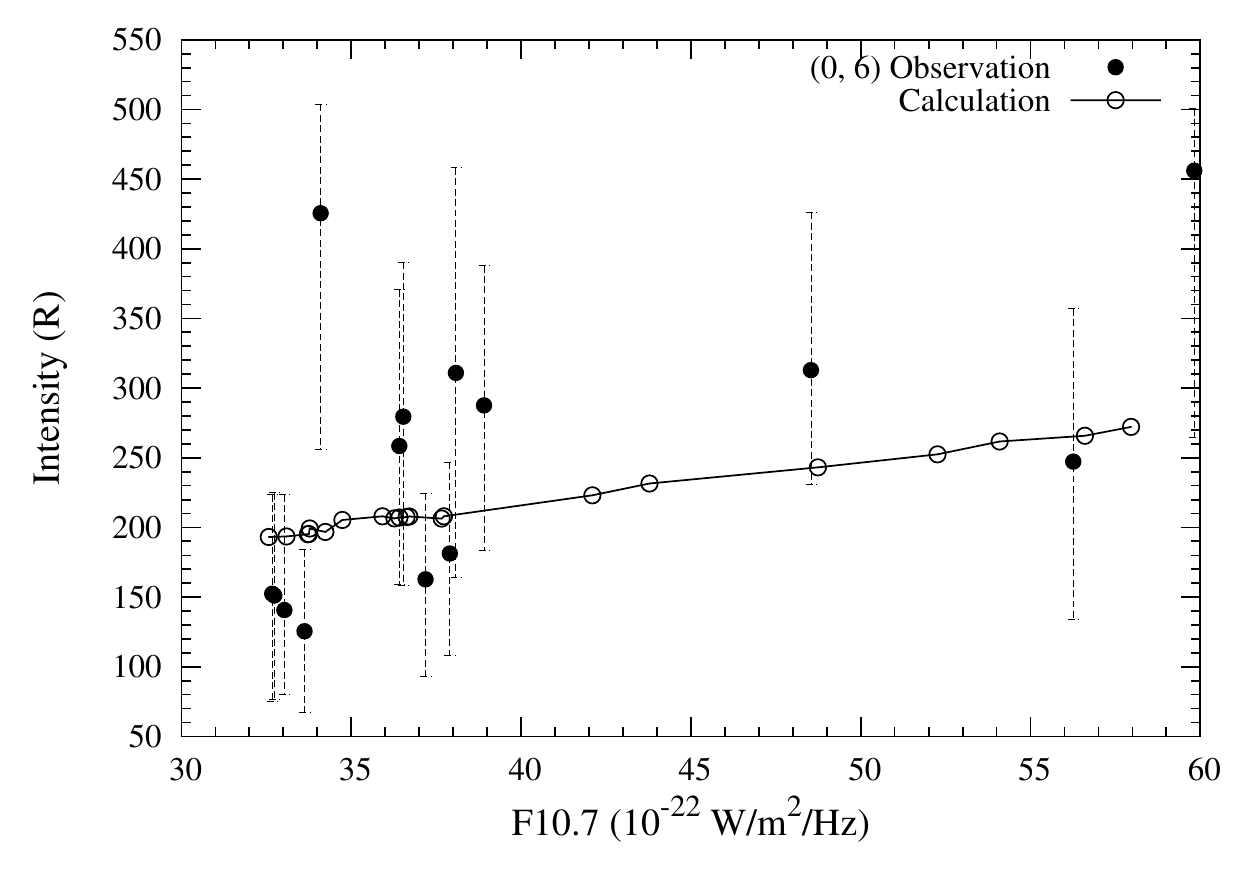}
\caption{Intensity variation of the \nt\ VK (0, 6) band 
with respect to solar index F10.7 ($W/m^2/Hz$) at Mars (scaled 
from the measured value at Earth). The observed intensity of the VK (0, 6) band
is taken from Figure 3 of \cite{Leblanc07}. The calculated intensity is
averaged-value between 120 and 170 km for the standard case
with \nt\ density in the atmosphere reduced by a factor of 3.}
\label{fig:n2flux}
\end{figure}

Another important model parameter, which affects the emission 
intensities is the solar EUV flux, whose variation is assumed to be
given by the F10.7 index. Solar EUV flux has been calculated 
using the F10.7 flux for the day of observation and scaled to the Mars 
according to its heliocentric distance. For the observations reported by
\cite{Leblanc07} the solar longitude of Mars varied between 101\dgr\ and 
171\dgr, which corresponds to change in the heliocentric distance of Mars
form 1.64 to 1.49 AU.
Figure~\ref{fig:n2flux} shows the variation of  VK (0, 6) band
intensity with respect to the F10.7 solar index at Mars. Calculations
are made for the standard case with the \nt\ density in the MTGCM model
reduced by a factor 3. Model calculated intensities are consistent with 
the observed values within the uncertainties of observation and model.

\section{Effect of various model parameters on Intensity}
\label{sec:ipi}
To evaluate the effect of various model input parameters, such 
as solar flux, cross sections, and model atmosphere, on the
VK band emissions, we have conducted a series of test
studies by changing one parameter at a time and compare the 
results with those of the standard case. The results are
presented in Table~\ref{tab:tableA} and discussed below.

\subsection{Electron impact cross sections for the triplet states}
\label{subsec:ele-xs}
Since electron impact on \nt\ is the source of excitation of 
forbidden triplet states of \nt, any change in electron impact 
cross sections will directly affect the VK band emission intensities.
Various measurements of the \nt\ triplet state cross sections 
were discussed in Section~\ref{sec:mi}. In the standard case
we have taken the recommended cross sections of \cite{Itikawa06},
which are fitted using the semiempirical relation given in
equation~(\ref{eq:jack}) (cf. Table~\ref{tab:table1} and Figure~\ref{fig:n2-xs}).
Instead of analytically fitted cross sections, if the triplet
state cross sections of \cite{Itikawa06} are used in the model, the
calculated triplet band intensities differ from the standard case
by less than 10\%.

Itikawa's recommended cross sections are based 
on the best values determined by the \cite{Brunger03}.
For the triplet states cross section, \cite{Brunger03} have 
estimated the uncertainty of the recommended cross sections as
$\pm$35\% ($ \pm $40\% at energies below 15 eV) for $ A^3\Sigma_u^+ $,
$\pm$35\% for $ B^3\Pi_g $ and $ W^3\Delta_u $,
$ \pm $40\% for $ B'^3\Sigma_u^- $, $ \pm$30\% for $ C^3\Pi_u $ and
$ \pm $40\% for $ E^3\Sigma_g^+ $ state. The integral cross sections (ICS)
of \cite{Johnson05} are derived from the differential cross sections 
(DCS) of  \cite{Khakoo05}. \cite{Johnson05} have
given the ICS at 8 energies between 10 and 100 eV,
with uncertainty for all states cross sections varying between $\pm$20\% to 
$\pm $22\%; at a few energy points it is as high as $\pm$35\%.

\begin{figure}    
\centering 
\includegraphics[width=30pc]{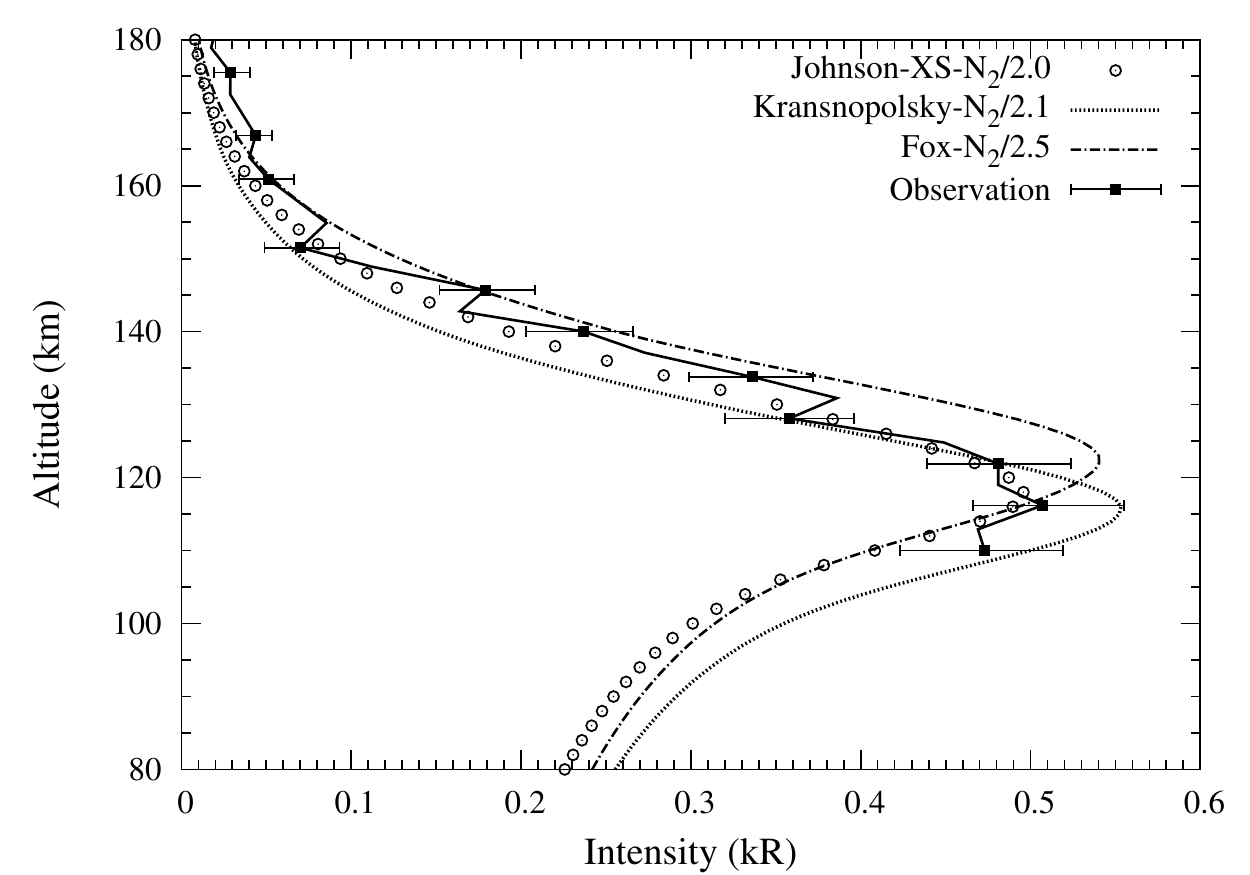}
\caption{The calculated limb intensity of the \nt\ VK (0, 6) band at SZA = 20\dgr.
The observed values are taken from \cite{Leblanc07}. The
calculated intensities are shown for the model atmospheres of \cite{Fox04}
(when the density of \nt\ is reduced by a factor of 2.5) and 
\cite{Krasnopolsky02} (the \nt\ density reduced by a factor of 2.1).
The intensity calculated by using the electron impact cross sections of 
\cite{Johnson05} is shown when the \nt\ density is reduced by a factor 
of 2.}
\label{fig:limb-cmp}
\end{figure}

To evaluate the effect of electron impact cross sections on the VK band emissions
we have taken two sets of cross sections; one from \cite{Cartwright77}, which were
renormalized by \cite{Trajmar83}, and second from the recent cross sections given
by \cite{Johnson05}.
The resulting VK band intensities are shown in Table~\ref{tab:tableA}.
The intensities calculated with the cross sections of \cite{Trajmar83} are
almost the same as in the standard case. However, when 
the cross sections of \cite{Johnson05} are used, the VK band 
intensities are reduces by 45\%, compared to the intensities computed for the
standard case, which is due to smaller cross sections of \cite{Johnson05}. The effect
of the smaller triplet state cross sections of \cite{Johnson05}
is  also seen on the limb intensities shown in the Figure~\ref{fig:limb-cmp}
where a reduction in \nt\ density  by a factor 2 is sufficient to fit the
SPICAM observed profile.
Thus, the electron impact triplet state excitation cross sections 
of \nt\ also help in constraining the \nt\ density in the model atmosphere.

\subsection{Input solar EUV flux model}
\label{subsec:flx-model}
SOLAR2000 model of \cite{Tobiska04} and EUVAC model 
of \cite{Richards94} are the two widely used solar flux models 
in the aeronomical calculations. In the standard case we have used
EUVAC model. To see the effect of input solar flux on
the VK emissions, we conducted a test study by taking the solar EUV flux 
from SOLAR2000 v.2.36 (S2K) model of \cite{Tobiska04} at
37 wavelength bins; the other input parameters remain the
same as in the standard case. The calculated integrated
overhead intensities are shown in the Table~\ref{tab:tableA}.
The calculated  intensities of VK bands using S2K model are $ \sim $15\%
larger than those calculated by using the EUVAC model.
This results in the requirement of a larger reduction in the
\nt\ density, that is, a factor of 3.4 compared to 3.0 for the
standard case to fit the observed limb profile of the VK (0, 6) band.

\subsection{Model atmosphere}
The importance of model atmosphere on the calculated intensities 
has been demonstrated in Section~\ref{subsec:loi}. The
\nt/\car\ ratio, which describes the abundance of molecular 
nitrogen in the atmosphere of Mars, is different in different 
model atmospheres. For the present study we have taken the
atmosphere from Bougher's MTGCM \citep{Bougher90,Bougher99,Bougher00} 
as used in study of \cite{Shematovich08} where the 
\nt/\car\ ratio is 2.8, 6.4 and 21\% at 120, 140 and 170 km, respectively.
\cite{Leblanc07} suggested that \nt/\car\ ratio is higher in 
the model atmosphere used by \cite{Fox79}. The recent models of 
\cite{Krasnopolsky02} are characterized by smaller abundances of \nt\
than that of \cite{Fox79}. The \nt/\car\ ratios are 2.6, 3.8, and 8.6\%
at 120, 140 and 170 km, respectively in \citeauthor{Krasnopolsky02}'s model.

We have used the model atmospheres of \cite{Krasnopolsky02} and 
\cite{Fox04} to study the effect of model atmosphere 
on the VK emission intensities. Figure~\ref{fig:limb-cmp} shows 
the calculated limb intensity of the VK (0, 6) band
for both model atmospheres at SZA 20\dgr\ (all other conditions are similar 
to the standard case). The emission peaks at $\sim $116 km in 
the case of \cite{Krasnopolsky02}, which is almost similar 
to  standard case ($\sim $118 km). But the emission peaks 
at higher altitude ($ \sim $123 km) when the model atmosphere of
\cite{Fox04} is used, which is due to higher \car\ abundance in 
her model. The intensities calculated using both models are
found to be larger than the observed values. To fit the observed limb
profile, the \nt\ density in the \cite{Krasnopolsky02} model has to be
reduced by a factor of 2.1, the \nt/\car\ ratios thus become
1.3, 1.8, and 4.4\% at 120, 140, and 170 km, respectively. In
the case of \cite{Fox04} model atmosphere, the required decrease in
\nt\ density is a factor of 2.5, which corresponds to the \nt/\car\ ratios
of 1.1, 1.9, and 5.3\% at 120, 140, and 170 km, respectively.

\subsection{Solar Cycle}
The solar cycle is approaching higher solar activity, and
MEX is currently orbiting Mars. We therefore hope to observe
the effects of higher solar activity on the Martian dayglow emissions.
Using the EUVAC model we have calculated the various \nt\ triplet band
emissions for high solar activity conditions similar to that of Mariner 6 
and 7 flybys when the  F10.7 index was $ \simeq $ 190 at 1 AU.
The model atmosphere for solar maximum conditions was taken from \cite{Fox04}; 
other model parameters are same as in the standard case.
The calculated height-integrated overhead intensities for
the VK bands are presented in  the Table~\ref{tab:tableA}, and those for
the other triplet bands in Table~\ref{tab:n2-oi}.
The calculated solar maximum intensities are larger by a factor of $\sim $1.5
than those of the standard case, and $ \sim $2.5 times larger than
those for Viking conditions.
As mentioned in section~\ref{subsec:loi}, the calculated and
observed limb profiles are consistent with each other when the
\nt\ density in the atmosphere is reduced by a factor of 3. 
If a similar situation prevails during high solar activity conditions,
then the calculated intensity of \nt\ VK band system would be smaller
by a factor of 2 to 3.

\section{Summary}
We have presented  models for the intensities of the \nt\ triplet
band systems in the Martian dayglow. We have used the analytical yield
spectra technique to calculate the steady state 
photoelectron flux, which in turn is used to calculate volume 
excitation rates of \nt\ VK bands and other triplet states.
The populations of various vibrational levels of the triplet states of \nt\ 
have been calculated considering direct excitation as well
as cascading from higher triplet states in statistical 
equilibrium conditions. Using calculated emission rates  the limb profiles
of the VK (0, 5), (0, 6), and (0, 7) bands have been calculated and 
compared with the SPICAM observed limb profile reported by \cite{Leblanc07}.
The observed and calculated limb profiles of the VK (0, 6) band are
in good agreement when the \nt\ density is reduced by a factor 
of 3 from those given by the  MTGCM model of
\cite{Bougher90,Bougher99,Bougher00}.
Overhead intensities of prominent transitions in 
VK, First Positive,  Second Positive, 
and $ W \rightarrow B $ bands have been calculated.

The effect of important model parameters, viz., electron impact 
\nt\ triplet state excitation cross sections, solar flux, solar activity, 
and model atmosphere, on emissions have been studied.
Changes in cross sections of \nt\ triplet states can alter the
calculated intensity by a factor of $ \sim $2. On the other hand,
the calculated intensities are $\sim $15\% larger when
the SOLAR2000 v.2.36 solar EUV flux model of \cite{Tobiska04} is used 
instead of the EUVAC model of \cite{Richards94}.
During high solar activity, when the F10.7 is similar to those at the times of
the Mariner 6 and 7 flybys, the calculated
intensities are about a factor of 2.5 larger than those
calculated for the low solar activity conditions of the Viking mission.
On using the model atmospheres of \cite{Fox04} and 
\cite{Krasnopolsky02}, a decrease in \nt\ density in their
atmospheric model by a factor of 2.5 and 2.1, respectively, is required to
reconcile the calculated VK (0, 6) band limb profile 
with the observed profile.

The most important parameter that governs the limb intensity of
VK band is the \nt/\car\ ratio. Constraining the \nt/\car\ ratio
by SPICAM observations, for different cases of model input 
parameters, we suggest that the \nt/\car\ ratio would be in the range of
1.1 to 1.4\% at 120 km, 1.8 to  3.2\% at 140 km, and 4 to 7\% 
at 170 km. Our study suggests that most of the atmospheric 
models  have \nt\ abundances that are larger than our derived values by factors
of 2 to 4. Clearly there is a need for improved  
understanding of the Martian atmosphere, and the SPICAM observations help
to constrain the \nt\ relative abundances. A decrease in the \nt\
densities in the atmospheric models, as suggested by our calculations,
would affect the chemistry and other aeronomical processes in
the Martian upper atmosphere and ionosphere.

\end{multicols}
%
%

%

%

\renewcommand{\thefootnote}{\fnsymbol{footnote}}

\newpage
\begin{center}
\begin{table}
\caption{Fitting parameters (equation~\ref{eq:jack}) for \nt\ triplet
state cross sections.}
\begin{tabular}{cccccccc}
\hline\noalign{\smallskip}
\multirow{2}{2.2cm} {Parameter} & \multicolumn{6}{c}{\nt\ states} \\
 \cline{2-7}\noalign{\smallskip}
  & A$^3\Sigma^+_u$ &  B$^3\Pi_g$ & C$^3\Pi_u$ & W$^3\Delta_u $ & B$'^3\Sigma^-_u$ 
 & E$^3\Sigma^+_g$\\
 \hline \\ [-0.4cm]
 Th\footnotemark[1]	& 6.17 & 7.35  & 11.03 & 7.36  & 8.16 & 11.9 \\[5pt]
 $\alpha $ & 1.00 & 3.00  & 3.20  & 1.50  & 1.70 & 1.70 \\[5pt]
 $\beta  $ & 1.55 & 2.33  & 1.00  & 2.30  & 1.50 & 3.00 \\[5pt]
 $\Omega $ & 2.13 & 2.50  & 2.70  & 2.60  & 2.12 & 3.00 \\[5pt]
 F         & 0.20 & 0.178 & 0.248 & 0.378 & 0.08 & 0.03 \\[5pt]
 W    	   & 6.99 & 7.50  & 11.05 & 8.50  & 8.99 & 12.0 \\[5pt]
 \hline
\end{tabular}

\footnotemark[1]{\small Threshold in eV.}
\label{tab:table1}
\end{table}
\end{center}

\begin{center}
\begin{table}
\caption{ \nt\ Vegard-Kaplan Band ($ A^3\Sigma^+_u \rightarrow X^1\Sigma^+_g $) 
height-integrated overhead intensity for different cases.}
\begin{tabular}{llccccccccc}
\hline
\multirow{3}{1.2cm}{Band $\nu'-\nu''$} & Band & \multicolumn{7}{c}{Overhead Intensity (R)}\\
\cline{3-9}
            & \multicolumn{1}{c}{Origin} & \multicolumn{1}{l}{Std.\footnotemark[1]}  & \multicolumn{1}{l}{$\rho$[\nt]} &\multicolumn{1}{l}{Viking} & \multicolumn{2}{l}{Cross section} &
            \multicolumn{1}{c}{Flux} &\multicolumn{1}{c}{Max.\footnotemark[5]}\\
\cline{6-7}
            & \multicolumn{1}{c}{(\AA)}	& \multicolumn{1}{l}{case}  & $/3.0$ & Cond. & CS-A\footnotemark[2] 
            & CS-B\footnotemark[3] & S2K\footnotemark[4] &  \\
\hline \\ [-0.4cm]
0-2		& 2216	& $1.5  $ & $0.5 $  & $0.9 $  & 1     & 1.4  & 1.7  & 2.3  \\[5pt]
0-3		& 2334	& $7.2  $ & $2.5 $  & $4.4 $  & 5     & 6.8  & 8.3  & 10.9 \\[5pt]
0-4		& 2463	& $19.4 $ & $6.8 $  & $11.7 $ & 13.3  & 18.3 & 22.2 & 29.3 \\[5pt]
0-5     & 2605 	& $34.3 $ & $12.1 $ & $20.7 $ & 23.5  & 32.4 & 39.4 & 51.8 \\[5pt]
0-6		& 2762	& $43.7 $ & $15.4 $ & $26.3 $ & 30    & 41.3 & 50.1 & 66.0 \\[5pt]
0-7 	& 2937	& $41.5 $ & $14.6 $ & $25.0 $ & 28.5  & 39.2 & 47.6 & 62.7 \\[5pt]
0-8		& 3133	& $30.7 $ & $10.8 $ & $18.5 $ & 21    & 29   & 35.2 & 46.4 \\[5pt]
0-9		& 3354	& $18   $ & $6.3  $ & $10.8 $ & 12.3  & 17   & 20.6 & 27.0 \\[5pt]
1-8		& 2998	& $25.9 $ & $9.1  $ & $15.5 $ & 18.8  & 25.3 & 29.6 & 38.4 \\[5pt]
1-9		& 3200	& $38   $ & $13.4 $ & $22.8 $ & 27.8  & 37.4 & 43.7 & 56.6 \\[5pt]
1-10	& 3427	& $35.9 $ & $12.7 $ & $21.5 $ & 26    & 35.1 & 41   & 53.2 \\[5pt]
1-11	& 3685	& $24   $ & $8.5  $ & $14.4 $ & 17.5  & 23.5 & 27.5 & 35.6 \\[5pt]
2-10	& 3270	& $12 $   & $4.2  $ & $ 7.1 $ & 8.8   & 11.9 & 13.6 & 17.3 \\[5pt]
2-11	& 3503	& $24.9 $ & $8.8  $ & $14.8 $ & 18.5  & 24.9 & 28.5 & 36.3 \\[5pt]
2-12	& 3769	& $26.9 $ & $9.5  $ & $16.0 $ & 20    & 26.9 & 30.8 & 39.3 \\[5pt]
2-13	& 4074	& $18.9 $ & $6.7  $ & $11.3 $ & 14    & 18.9 & 21.7 & 27.6 \\[5pt]
3-13	& 3857	& $16.3 $ & $5.7  $ & $9.7  $ & 12.3  & 16.5 & 18.7 & 24.0 \\[5pt]
3-14	& 4171	& $18.1 $ & $6.4  $ & $10.8 $ & 13.7  & 18.3 & 20.7 & 26.7 \\[5pt]
\hline
\end{tabular}

\footnotemark[1]{\small Standard case. See text for details} \\
\footnotemark[2]{\small Cross sections taken from \cite{Johnson05}.}\\ 
\footnotemark[3]{\small Cross sections taken from \cite{Trajmar83}.}\\ 
\footnotemark[4]{\small SOLAR2000 model of \cite{Tobiska04}.}\\ 
\footnotemark[5]{\small Solar maximum flux for condition similar to Mariner 6 flyby 
(F10.7 $ \simeq $ 190).}
\label{tab:tableA}
\end{table}
\end{center}

\begin{center}
\begin{table}
\caption{Calculated height-integrated overhead intensity of 
\nt\ triplet emissions.}
\begin{tabular}{cccc}
\hline
Band & Band Origin & \multicolumn{2}{c}{Intensity (R)} \\
\cline{3-4}
($ \nu' - \nu''$) & \AA\ & Std.\footnotemark[1] & Max.\footnotemark[2] \\
\hline\noalign{\smallskip}
\multicolumn{4}{c}{First Positive  
$B^3\Pi_g$-- $A^3\Sigma^+_u$}\\[5pt]
0-0	&	10469	& 	60.9	&	95.4 \\
0-1	&	12317 	& 	32.7   	&	51.2 \\
0-2 &	14895	&	9.8  	&	15.3 \\
1-0	&	8883	&	96  	&	149.7\\
1-2	&	11878	&	17.8  	&	27.8 \\
1-3	&	14201	&	13.6  	&	21.2 \\
2-0	&	7732	&	46.9	&	72.9 \\
2-1	&	8695	&	60.6	&	94.2 \\
2-2	&	9905	&	11.2 	&	17.5 \\
3-1	&	7606	&	64.3 	&	100  \\
3-2	&	8516	&	16.8  	&	26.1 \\
3-3	&	9648	&	20.3 	&	31.5 \\
4-1	&	6772	&	19.6 	&	30.5 \\
4-2	&	7484	&	49.2  	&	76.4 \\
4-4	&	9404	&	14.8  	&	22.9 \\
5-2	&	6689	&	22.1  	&	34.4 \\
5-3	&	7368	&	26.2  	&	40.7 \\
6-3	&	6608	&	18  	&	28   \\
7-4	&	6530	&	12  	&	18.5 \\[10pt]
\multicolumn{4}{c}{Second Positive  
$C^3\Pi_u$-- $B^3\Pi_g$} \\[5pt]
0-0	&	3370	&	32.1	&	51   \\
0-1	&	3576	&	21.7	&	34.5 \\
0-2	&	3804	&	8.7		&	13.9 \\
1-0	&	3158	&	8.3	    &	13.2 \\ [10pt]
\multicolumn{4}{c}{Wu-Benesch ($W^3\Delta_u$-- $B^3\Pi_g$)} \\[5pt]
2-0	&	33206	&	4   	&	6.4 \\
3-0	&	22505	&	3.4 	&	5.4 \\
3-1	&	36522	&	3   	&	4.7 \\
4-1	&	24124	&	5.1 	&	8   \\
5-1	&	18090	&	4.7 	&	7.4 \\
5-2	&	25962	&	4.5 	&	7.0 \\
6-2	&	19193	&	5.8 	&	9.0 \\
7-2	&	15281	&	4.7 	&	7.3 \\
7-3	&	20421	&	5.1 	&	8.0 \\
8-3	&	16112	&	5.1 	&	8.0 \\
9-4	&	17024	&	4.4 	&	6.9 \\
\hline
\end{tabular}
\\ \footnotemark[1]{\small Standared Case. See text for details}\\
\footnotemark[2]{\small Solar maximum flux for condition similar to Mariner 6 flyby.}
\label{tab:n2-oi}
\end{table}
\end{center}

\clearpage

\end{document}